\documentclass[iop]{emulateapj}
\usepackage{color}
\usepackage{natbib}
\definecolor{lila}{rgb}{0.3,0,0.7}
\definecolor{grau}{rgb}{0.5,0.5,0.5}

\newcommand{\Teff}{T_{\rm eff}}
\newcommand{\Teq}{T_{\rm eq}}
\newcommand{\Tint}{T_{\rm int}}
\newcommand{\Leff}{L_{\rm eff}}
\newcommand{\Leq}{L_{\rm eq}}
\newcommand{\Eint}{E_{\rm int}}
\newcommand{\ME}{M_{\oplus}}
\newcommand{\RE}{R_{\oplus}}

\newcommand{\fig}{Fig.$\:$}
\newcommand{\tab}{Table$\:$}
\newcommand{\sect}{\S$\:$}

\shorttitle{Thermal evolution and structure of GJ$\:$1214b}
\shortauthors{Nettelmann et al.}

\begin{document}

\title{Thermal evolution and structure models of the transiting super-Earth GJ$\:$1214b}
\author{N.~Nettelmann\footnotemark[1,2], J.~J.~Fortney\footnotemark[1,3],
U.~Kramm\footnotemark[2] and R.~Redmer\footnotemark[2]}

\footnotetext[1]{Department of Astronomy and Astrophysics, University of California, Santa Cruz, CA 95064}
\footnotetext[2]{Institut f\"ur Physik, Universit\"at Rostock, D-18051 Rostock} 
\footnotetext[3]{Alfred P.~Sloan Research Fellow}

\slugcomment{Accepted to ApJ, 03/2011}

\begin{abstract}
The planet GJ 1214b is the second known super-Earth with a measured mass and radius. 
Orbiting a quiet M-star, it receives considerably less mass-loss driving X-ray and UV 
radiation than CoRoT-7b, so that the interior may be quite dissimilar in composition, 
including the possibility of a large fraction of water. We model the interior of GJ$\:$1214b 
assuming a two-layer (envelope+rock core) structure where the envelope material is either 
H/He, pure water, or a mixture of H/He and H$_2$O.  Within this framework we perform models 
of the thermal evolution and contraction of the planet.  We discuss possible compositions 
that are consistent with $M_p=6.55\:\ME$, $R_p=2.678\:\RE$, an age $\tau=3-10$~Gyr, and the 
irradiation level of the atmosphere.  These conditions require that if water exists in the 
interior, it must remain in a fluid state, with important consequences for magnetic field 
generation.  These conditions also require the atmosphere to have a deep isothermal region 
extending down to 80$-$800~bar, depending on composition. Our results bolster the suggestion 
of a metal-enriched H/He atmosphere for the planet, as we find water-world models that lack 
an H/He atmosphere to  require an implausibly large water-to-rock ratio of more than 6:1.  
We instead favor a H/He/H$_2$O envelope with high water mass fraction ($\sim$~0.5$-$0.85), 
similar to recent models of the deep envelope of Uranus and Neptune. Even with these high 
water mass fractions in the H/He envelope, generally the bulk composition of the planet can 
have subsolar water:rock ratios. Dry, water-enriched, and pure water envelope models differ 
to an observationally significant level in their tidal Love numbers $k_2$ of respectively 
$\sim 0.018$, $\sim 0.15$, and $\sim 0.7$. 
\end{abstract}


\section{Introduction}

Among the nearly 500 detected planet candidates, and in particular among the $\sim 80$ 
planets for which both the mass and radius have been determined, our Earth is the 
only planet that is known to harbor liquid water oceans on a solid surface crust. 
These conditions have proven favorable for the development of life forms.
With the discovery of CoRoT-7b ($M_p=4.8\pm 0.8\:\ME$; \citealt{Leger+09}) and 
GJ$\:$1214b ($M_p=6.55\pm 0.98\:\ME$; \citealt{Charb+09}),  the search for other 
habitable worlds has just recently passed the milestone of discovering transiting 
extrasolar planets in the 2-10 $\ME$ super-Earth mass regime.

Higher-mass planets such as Uranus ($M_p=14.5\:\ME$) are predicted by interior models 
to retain a H/He-rich atmosphere \citep{Hubb+95} whose size may vary depending 
on $M_p$, $R_p$ and temperature.  Pressures at the bottom of this envelope are high, 
and may reach 1~to~1000~kbar if the planet is of Neptune-size \citep{N-GJ436b+10} 
or even several Mbar if of Saturn-size \citep{Guillot99} before a presumably solid core 
is reached. Lower-mass objects on the other hand such as Mars or Ganymede have been 
observed not to retain a thick enough atmosphere that could prevent the planet's surface 
from cooling below the freezing point of water. 

CoRoT-7b --orbiting a Sun-like star at short orbital distance $a_p=0.017$~AU--
falls into the intermediate super-Earth mass regime; the stellar extreme ultraviolet 
(EUV) flux it receives is so strong that its current atmosphere is either a tiny remnant 
of an initially massive gaseous envelope, or hot evaporating core material at an 
equilibrium temperature $\Teq=1800-2600$~K \citep{Valencia+10,Jackson+10}. 
In contrast, GJ$\:$1214b ($a_p=0.0144\:$AU) orbits an M star of $\sim 3\times 10^{-3}$ 
smaller luminosity \citep{Charb+09} translating into a planet-average $\Teq\leq 555$~K, 
only a factor of two higher than that of Earth. Hence the discovery of GJ$\:$1214b 
manifests an important step toward a detection of an extrasolar ocean planet. 

\citet{RogSea10} investigated the response of interior models to the uncertainties in 
$M_p$, $R_p$, and intrinsic temperature $\Tint$ and showed that GJ$\:$1214b might have 
a $(10^{-4}-0.068)\times M_p$ thick H/He atmosphere, or else an outgassed H atmosphere,
or a water envelope atop a silicate-iron core with an ice:rock (I:R) ratio of 0.06 to $\infty$.

In this paper we adopt the fiducial $\{M_p, R_p\}$ values and investigate how the 
unknown temperatures of the deep interior can be constrained by thermal evolution calculations. 
Our models are two-layer models with one homogeneous envelope overlying a rock core. 
We take into account mass loss during evolution and explore how that affects the 
possible mass of an outer H/He layer (\sect\ref{ssec:res_class1}). 
In \sect\ref{ssec:res_class2} we consider pure water atmospheres and ask whether 
condensation or even solidification of water could then have occured within 10 billion 
years of cooling. In line with recent transmission spectrum measurements that indicate 
70\% or more water by mass in the atmosphere  \citep{Bean+10}, we vary in
\sect\ref{ssec:res_class3} the envelope water mass fraction between 50 and 100\% and
suggest plausible models with about $0.2\times$solar I:R ratio.  
In \sect\ref{sec:discuss} we discuss our model assumptions and propose to discriminate 
between our three classes of models (dry, water, water-rich envelope) by observationally 
determining the Love number $k_2$ and the mean molecular weight of the atmosphere. 
Our method of modeling this planet is explained in \sect\ref{sec:methods}, where we 
describe the irradiated atmosphere grid (\sect\ref{ssec:atm}), applied equations of 
state (\sect\ref{ssec:eos}), mass loss (\sect\ref{ssec:massloss}), and the calculation 
of structure and evolution (\sect\ref{ssec:StrukEvol}).

\section{Methods}\label{sec:methods}

In this section we describe the four components that our interior models rely on: 
the model atmosphere grid, the equations of state used, structure assumptions, and 
the thermal evolution to the present state.  The evolutionary models have some 
similarities to what has previously been applied to hot Neptunes and hot Jupiters 
\citep[e.g.][]{Fortney+07,Baraffe+08}, but with additional complications due to 
radiogenic heating.  As the planet's interior cools, the external radiative zone 
grows deeper~\citep{Guillot+96}, reaching a depth of up to several hundreds of bars.  
The transition pressure of the atmosphere from radiative to adiabatic, at the current 
time, $P_{ad}(t_0)$, is a quantity we aim to constrain with our evolution model.  
We also perform explorations of the planet's structure as a function of $P_{ad}(t_0)$, 
to investigate the full range of hotter, higher entropy interiors (lower $P_{ad}(t_0)$) 
and cooler, lower entropy interiors (higher $P_{ad}(t_0)$) that may be possible today.

\subsection{The model atmosphere}\label{ssec:atm}

Under the assumption the deep envelope layers convect efficiently, it is the 
radiative atmosphere atop the convective region that serves as the bottleneck 
for interior cooling, just as in Neptune, Uranus, and giant planets generally 
(e.g., \citealt{Hubbard77}).  
Observations of GJ1214b's atmosphere are consistent with a water-dominated composition
as well as with a H/He atmosphere with clouds or hazes \citep{Bean+10}. Given our 
current ignorance of the composition of the atmosphere, we consider the two likely 
end-member cases, either a H/He-dominated atmosphere, or a pure steam atmosphere,
and find that the opacity is dominated by water vapor in either case \citep{MillerRicciFor10}. 
For planetary structure, we assume chemical equilibrium in the atmosphere, 
thereby ignoring possible alterations of the T-P profile through photoionization.

For planetary evolution, a grid of model atmospheres is generally used as the 
upper boundary condition, see, for instance, \citet{Fortney+07}. 
These grids relate the specific entropy ($s$) of the convective
interior, surface gravity ($g$) of the planet, and the intrinsic
effective temperature ($\Tint$) from the interior.  We have
computed such a grid from $\Tint=175$ K down to 30 K, with the correct limiting
behavior down to 0 K (an exhausted interior) across surface gravities
from 100 to 1500 cm s$^{-2}$.  The grid is computed at 50$\times$ solar metallicity, 
under the assumption of complete redistribution of absorbed stellar flux  
(meaning f=1/4, see \citealt{Fortney+07, MillerRicciFor10}), and no clouds.  Similar models, 
which describe the technique in more detail, are found in \citet{Fortney+07}.  
The opacity database is described in \citet{Freedman+08}.  We note that this very high 
metallicity is realistic given that Neptune and Uranus are 30-60 times solar in carbon 
(see \citet{GuiGau09}, for a review).

We use this grid for all evolution calculations of our models, whether they posses thin 
H/He - atmospheres or pure steam atmospheres. This is certainly a broad brush 
treatment for a wide range of possible atmospheres, but given our current ignorance 
regarding the planet's atmosphere, we feel our treatment is justified.
The importance and utility of the coupled model atmosphere/interior cooling calculation 
is that it allows us to estimate $\Tint$ and $P_{ad}$ as a function of time.  
For instance, in the recent work of \citet{RogSea10}, the value of $\Tint$ 
was not calculated, but was extrapolated from evolutionary models of \citet{Baraffe+08}, 
for higher mass objects.  Generally, we find a 15~K lower $\Tint$, meaning a colder 
interior, than \citet{RogSea10} used. 
While $\Tint$ may change from 175 down to 30~K during evolution, we find that the 
effective temperature $\Teff$ remains nearly constant within 562 to 557~K, which is 
close to the zero-albedo, planet-average equilibrium temperature of 555~K.

\subsection{Equations of state}\label{ssec:eos}

Metal-rich\footnote{The label \emph{metals} comprises all elements heavier than H and He.} 
planets such as super-Earths are generally suspected to harbour a variety of materials. 
We aim to represent this variety in a simplified manner 
by confining silicates and iron into a 'rocky' core, and H, He, and water to an envelope.  
For core material we use the $P-{\rho}$ relation for rocks by \citet{HM89} which describes
an adiabatic mixture around $10^4$~K of 38\%~SiO$_2$, 25\%~MgO, 25\%~FeS, and 12\%FeO.
Such kind of rocks' mass fraction of Si, Mg, and Fe is, respectively, about 0.5, 0.62, 
and 1.05 times that of the bulk Earth \citep{McDonSun95}.
For H/He envelopes we use the interpolated hydrogen and helium EOS 
developed by \citet{SCvH95}. For water we use H$_2$O-REOS, which was applied to Jupiter 
\citep{N-Jupiter+08}, Uranus and Neptune \citep{FortneyINGM10} and in a slightly modified 
version to CoRoT-7b \citep{Valencia+10}. This water EOS comprises various water 
EOS appropriate for different pressure-temperature regimes. It includes the melting curve 
and phase Ice I \citep{FeiWag06}, the saturation curve and liquid water \citep{WagPru02}, 
vapor and supercritical molecular water (SESAME 7150, \citealp{SESAME}), and for 
$T\geq 1000$ K and $\rho\geq 2$ g cm$^{-3}$ supercritical molecular water, ionic water, 
superionic water, plasma, ice XII, and ice X based on FT-DFT-MD simulations 
\citep{French+09}. At pressures below 0.1~GPa, and/or temperatures below 1000~K, H$_2$O-REOS 
relies on  Sesame~EOS~7150 \citep{SESAME}.
For mixtures of hydrogen, helium, and water we use H-REOS for hydrogen \citep{N-Jupiter+08}, 
H$_2$O-REOS for water, and an improved version \citep{KerleyHe} of the helium Sesame 
EOS 5761 \citep{SESAME}. Other materials are not considered here 
\footnote{ 
We do not consider the lighter and more volatile ices CH$_4$ and NH$_3$. We performed
simple tests by perturbing warm H$_2$O-REOS adiabats with the zero-temperature $P-\rho$ 
relations for CH$_4$ and NH$_3$ by \citet{ZT78}.  In a solar C:N:O proportion, 
this reduces the density by $\sim 0.5$ g cm$^{-3}$ if $P>20$ GPa, but enhances the density 
for smaller pressures, due to neglection of finite temperature effects which are important 
for supercritical ices.  Since these EOS are not of comparable quality to those used for 
all other components, we do not include them here.}.

\subsection{Mass loss}\label{ssec:massloss}

Mass escape caused by stellar energy input is known to occur from the highly irradiated 
atmosphere of the hot Jupiter HD 209458b \citep{VidalMadjar+03,Yelle04,Erkaev+07,MurrayClay09}, 
and shown to have a significant impact on the current composition of super-Earth CoRoT-7b 
\citep{Valencia+10, Jackson+10}. While the sun-like star CoRoT-7 irradiates the planet with 
a present X-ray and ultra-violet energy flux (XUV)  
$\rm F_{XUV}=5\times 10^5\:erg\,cm^{-2}\,s^{-1}$ \citep{Valencia+10}, GJ$\:$1214 is supposed 
to be an inactive M star \citep{Charb+09}. Assuming it obeys the empirical relation for the 
surface energy flux of M stars, $F_{\star,\rm XUV}/F_{\star,\rm bol}=(1-100)\times 10^{-5}$ 
\citep{Scalo+07}, then with $L_\star=0.0033L_\sun$, $R_\star=0.211 R_\sun$, 
the energy flux $F_{\rm XUV}=F_{\star,\rm XUV}(R_p^2/4a_p^2)$ received by GJ$\:$1214b is 
only (0.8$-$80)~erg~cm$^{-2}~s^{-1}$, i.e.~$(0.16-16)\times 10^{-4}$ that of CoRoT-7b, at the 
current time. We then expect XUV irradation to have a comparatively lesser, but still 
important, influence on the atmospheric mass through time.

We use the energy limited escape model of \citet{Erkaev+07} to investigate the mass-loss 
history of the planet.  With a heat absorption efficiency $\varepsilon=0.4$, and a correction 
factor $K_{\rm tide}=0.95$ accounting for the height decrease of the Roche-lobe boundary 
through tidal effects \citep{Erkaev+07}, the energy-limited mass escape 
rate $\dot{M}_{\rm esc} = (\varepsilon\, F_{\rm XUV} R_p^3)\,(G\,M_p\,K_{\rm tide})^{-1}$
\citep[following][]{Valencia+10} is $(0.012-2.8)\times 10^8$~g~s$^{-1}$.  This is 
$\approx(0.02-6)\times 10^{-3}$ the mass loss of present (rocky) CoRoT-7b, with a value 
of $(1-100)\times 1.84\times 10^6$~g~s$^{-1}$, for the fiducial GJ$\:$1214b values 
$M_p=6.55\:\ME$ and $R_p=2.678\RE$.  
Correcting $M_{\rm esc}\sim r_1^2 R_p$ for the altitude $r_1>R_p$ where the XUV flux is 
absorbed \citep{Lammer+03}, the actual value can further rise by a factor of 10.
Within 1~Gyr, this mass loss accummulates to $(1-100)\times 10^{-4}\:\ME$. 
For a low-mass atmosphere of only about 1\% of GJ$\:$1214b's total mass (which is quite 
possible for a small mean molecular weight atmosphere: see \sect\ref{ssec:res_class1}), 
the fraction of the atmosphere lost during the 3 to 10 Gyr lifetime of this planet can be 
large enough to influence its cooling behavior.  In particular, according to the estimates 
above and assuming constant mass loss over time, the initial atmosphere can have been larger 
by $(1-100)\times(0.5-1.5)\%$. We therefore must include mass loss in our evolution calculations 
of H/He envelope models, as described below.

\subsection{Interior structure and evolution modeling}\label{ssec:StrukEvol}

\subparagraph{Three classes.}
We consider three classes of hydrostatic two-layer interior models of present 
GJ$\:$1214b assuming a homogeneous envelope above a rocky core. The three classes 
differ in the materials constituting the envelope. Class I models have a H/He envelope
with solar He mass fraction $Y=M_{\rm He}/(M_{\rm H}+M_{\rm He})=0.27$. Class {\rm II} 
models have a pure water envelope (``water worlds''), and class III models a H/He/H$_2$O 
envelope with variable water mass fraction $Z_1$, and also $Y=0.27$. 
The rock core mass of present GJ$\:$1214b models is found by the condition to match the 
radius $R_p=2.678\RE$ (recently confirmed through both optical and infrared 
photometry~\citep{Sada+10}) for a given planet mass $M_p=6.55\ME$ and surface thermal 
boundary condition. We do not consider different possible $R_p-M_p$ pairs within the 
observational error bars $\sigma_{M_p}=\pm 1.0\ME$ and $\sigma_{R_p}=\pm 0.13\RE$ as such 
work has already been presented by~\citet{RogSea10}.  For the thermal structure of the 
atmosphere we apply the solar composition model atmosphere between 20~mbar and 10~bar to 
model classes I and III, and to model class II the water model atmosphere between 20~mbar 
and 1~bar from~\citet{MillerRicciFor10}.  At higher pressure in the radiative atmosphere,
we assume an isothermal temperature---a reasonable assumption \citep{MillerRicciFor10} 
and given and our general understanding of highly irradiated atmospheres.

\subparagraph{Calculating the structure.}
We choose 20~mbar as the low-pressure boundary of our models.  This is a choice 
of convenience, since our water EOS ends at this pressure, but it is also realistic.  
The wide-band optical transit radius for the planet is at $\sim$10~mbar (E.~Miller-Ricci, 
personal communication).  This is consistent with the cloud-free atmosphere calculations 
of~\citet{Fortney+03} for HD$\:$209458b, as well.

The high-pressure boundaries of the model atmospheres of respectively 1 and 10~bar are 
chosen within the isothermal part of the atmosphere, before it transitions to the adiabatic 
interior at some pressure $P_{ad}$. For present time ($t_0$) structure models, we consider 
the transition pressure $P_{ad}(t_0)$ a variable parameter and investigate the response 
of the core mass and the cooling time on the choice of $P_{ad}(t_0)$.

Since the model atmospheres predict almost constant equilibrium abundances of H, He, 
and H$_2$O, we derive the mass density $\rho(P,T(P))$ in the atmosphere $T(P)$ from an 
EOS table for constant composition. Given $M_p$, $R_p$, and the $P-\rho$ relations 
according to the EOS $\rho(P,T(P))$ in the atmosphere, and constant $s(P,T,\rho(P,T))$ 
in the adiabatic interior, we obtain internal profiles $m(r)$, $T(r)$, and $P(r)$ by 
integrating the equation of hydrostatic equilibrium $dP/dr=-Gm\rho/r^2$ from the surface 
toward the center. Mass conservation $M_p=\int_0^{R_p} dr\: 4\pi r^2\:\rho(r)$ is ensured 
by the proper choice of the rock core mass $M_{core}$.

\subparagraph{The rock core assumption.}
We assume the rocky core to be appropriatly described by an EOS of \emph{homogeneous, 
adiabatic} 'rocks' at all times (compare Rogers and Seager 2010: differentiation into 
an iron core and a silicate layer of $\rm Mg_{0,1}Fe_{0,1}SiO_3$ at uniform temperature).  
Differentiation will have occurred when temperatures in the primitive rock core rose 
above the melting temperature of iron during formation, and will have affected the thermal 
evolution. In the Earth, solidification of the inner iron core still causes a bouyancy of 
light elements driving convection of the outer iron core, and supports --together with 
gravitational energy release from core shrinking-- subsolidus convection of the silicate 
mantle. If the melting line of iron rises steeply with temperature as indicated by ab 
initio data~\citep[see][for an overview]{Valencia+10}, also the central part of a several 
earth mass core of GJ$\:$1214b might transition from liquid to solid iron due to high 
pressure up to $\sim 15$ Mbar, supporting the assumption of an adiabatic interior.

Due to the poorly constrained iron mass fraction of GJ$\:$1214b and uncertainties 
in the iron melting line, the deep interior could potentially be fully liquid or 
solid and isothermal. Since the equations of state of rocky materials at high pressure 
above few Mbar are not well known and the effect of temperature on the $P-\rho$ relation 
is negligible \citep{Seager+07}, we believe the $P-\rho$ relation of the rock-EOS used 
is appropriate for our purpose of determining the core mass and its contribution to the 
cooling time. We denote by $T_{core}$ the temperature at the core-mantle boundary and
assume its time derivative $T_{core}/dt$ to be representative for the whole core.
 
\subparagraph{Calculating the evolution.}
Chosing $P_{ad}(t_0)$ for class I and II models, and $P_{ad}(t_0)$ and $Z_1$ for class 
III models uniquely defines the core mass of resulting interior models. A selection of six 
such present time interior models is shown in \tab\ref{tab:SM}. For class I and II models, 
we calculate the cooling curve by first generating $\sim 50$ profiles with decreasing 
transition pressures $P_{ad}(t_0)\geq P_{ad}>20$~mbar values, thereby increasingly warmer 
interiors. For each of these intermediate profiles, core mass and composition are 
conserved. As the interior becomes warmer with decreasing $P_{ad}$, the planet radius 
$R_p$ rises. In order to obtain the cooling curve $R_p(t)$, we integrate the energy 
balance equation 
\begin{equation}\label{eq:leffleqlint}
	\Leff-\Leq = -\frac{d\Eint}{dt}
\end{equation}
backward in time, starting with the present time structure models. In 
Eq.(\ref{eq:leffleqlint}), $\Leff=4 \pi R_p^2 \sigma\Teff^4$ is the net luminosity 
the planet radiates into space, and $\Leq=4\pi R_p^2 \sigma\Teq^4$ is the 
stellar energy absorbed. The difference $\Leff-\Leq\sim \Tint^4$ is provided by 
our atmosphere grid $\Tint(g,s)$, see \sect\ref{ssec:atm}, and sets the intrinsic 
luminosity the planet can radiate away from the interior through its atmosphere 
at a given gravity and internal entropy. Given $\Tint$, we can then derive the
time interval $dt$ it needs to lose the intrinsic energy $d\Eint$,
\begin{equation}\label{eq:lint}
	\frac{d\Eint}{dt} = \int_{M_{core}}^{M_{p}}\hspace*{-3mm} dm\: \frac{T\,ds}{dt} 
            + c_v\,M_{core}\frac{dT_{core}}{dt} - L_{radio}.
\end{equation}
Expression (\ref{eq:lint}) accounts for the heat loss $\delta q=T(m)ds$ of each envelope 
mass shell $dm$, the heat loss of the core due to cooling, and the energy gain 
$L_{radio}$ of rocky core material due to decay of radioactive elements (see below). 

Experimental data for the specific heat $c_v$ of warm, compressed rocks at $P>2\:$Mbar 
are not available. Ab initio calculations for iron at a few Mbar and several thousand K 
(Earth's core conditions) predict $c_v\approx 0.5\:\rm J\,K^{-1}g^{-1}$ \citep{Alfe+01}, 
while $c_v=1.0\:\rm J\,K^{-1}g^{-1}$ was formerly applied by \citet{Guillot+95} to the 
core of Jupiter and by \citet{Valencia+10} to the silicate-iron interior of CoRoT-7b.
We aim to bracket the uncertainty in $c_v$ by chosing $c_v=0.5-1.0\:\rm J\,K^{-1}g^{-1}$. 
For given $\Tint$, the time interval required to lose $\Eint$ rises with $c_v$. 

\subparagraph{Radiogenic heat.}
As we will see, our class I and III planet models have large rocky cores. Modeling 
the radiogenic heat from the rocky portion of the interior is important to 
accurately calculate the thermal evolution.  
For $L_{radio}$ we consider the isotopes $^{238}$U, $^{235}$U, $^{232}$Th, and 
$^{40}$K. We adopt isotopic abundances and element abundances of meteorites for 
the elements U, Th, K, and Si\footnote{The Si mass abundance of meteorites (10.65\%)
is unequal that of the rock EOS (16.8\%).} as given in \citet{AndGrev89}. 
With respective half-life times in Gyrs of 4.468, 0.704, 14.05, and 1.27, 
and respective decay energies in MeV of 4.27, 4.679, 4.083, and 1.33\footnote{These 
values are in part an average over the surprisingly wide spread of literature values, 
see e.g.~http://ie.lbl.gov/education/,  and in part over decay chains ($^{40}$K)}, 
we find a radioactive energy release for $1\ME$ of meteoric material on Earth today of 
$2.3\times10^{13}\:{\rm J\,s^{-1}} /1\ME$ $(=:L_{radio,meteor}(t_0))$ and 4.56~Gyr ago 
of $2.7\times10^{14}\:{\rm J\,s^{-1}}/1\ME$ $(=:L_{radio,meteor}(0))$. The dominant 
contribution during this time interval is mostly due to $\beta^-$ decay of $^{40}$K 
into $^{40}$Ca. Extension of the radioactive decay law to 10 Gyr ago would increase 
$L_{radio}$ to $20\times L_{radio,meteor}(0)$, an unrealistically high value for rock 
material in the young universe. 
Of course, for small core masses ($0.1M_p$), cooling of the core does not 
significantly affect the model's cooling time, but for large core mass models it does, 
and in particular the choice of the initial value $L_{radio}(0)$ matters. 
Therefore, we define a cosmological luminosity $L_{cosmo}:=L_{radio,meteor}(0)$ that we 
require each cooling model of GJ$\:$1214b to start with, within 10\%. Models with long 
cooling times (10~Gyr) will then have $L_{radio}(t_0) \approx 0.1\times L_{radio,meteor}(t_0)$, 
and models with short cooling times (3~Gyr) will have 
$L_{radio}(t_0)\approx 2\times L_{radio,meteor}(t_0)$.
By this choice of $L_{cosmo}$ we avoid extremely high or low initial values.  


\begin{deluxetable*}{ccccccccc}
\tabletypesize{\scriptsize}
\tablecaption{GJ$\:$1214b structure model input and output data\label{tab:SM}}
\tablewidth{0pt}
\tablehead{
  \colhead{Label} & \colhead{$M_{core}$} & \colhead{$P_{ad}(t_0)$} & \colhead{$Z_1$} 
  & \colhead{envelope material} & \colhead{$\dot{M}$} & \colhead{$k_2$} 
	& \colhead{$T_{\rm int}$} & age\\
  & ($\ME$) & (bar) & & & ($10^7$ g/s)& & (K) & (Gyr)   
}
\startdata
Ia & 6.464  & 300  & 0    & H/He        & 1.84 & 0.0183 & 42.7 & 3.1 \\
Ib & 6.434  & 800  & 0    & H/He        & 1.84 & 0.0170 & 31.8 & 7.2 \\
IIa & 0.873  &  80  & 1    & water       &-     & 0.5769 & 51.9 & 3.05 - 3.16\\
IIb & 0.203  & 300  & 1    & water       &-     & 0.737  & 37.6 & 9.23 - 9.26\\
IIIa & 4.432  & 120  & 0.85 & H/He/water  &-     & 0.14336 & 48.4 & 2.92 - 3.22 \\
IIIb & 4.016  & 400  & 0.85 & H/He/water  &-     & 0.186  & 35.6 & 9.21 - 10.10
\enddata
\tablecomments{These structure models have $R_p=2.678\RE$ and $M_p=6.55\ME$.} 
\end{deluxetable*}

\subparagraph{Evolution with mass loss.}
For a given planet radius, low mean molecular weight atmospheres or envelopes are 
also of low mass, and hence such envelopes may lose a larger relative mass fraction 
than large mean molecular weight atmospheres.  We take into account mass loss only 
for the H/He envelope models, e.g.~models Ia,b in \tab\ref{tab:SM}. For each neighbored 
pair of interior profiles as given by $P_{ad}$, we calculate the mass $dM$ lost during 
a time interval $dt$ self-consistently using a Newton-Raphson scheme to find the root 
of the function $f(dt(dM))=dM-\dot{M}_{esc}dt(dM)=0$.

\subparagraph{Other contributions.}
We did not include the effect of tidal heating due to tides raised on the planet 
by the star. This additional amount of energy would tend to delay the cooling time, 
in particular in the past when the eccentricity $e$, and hence the tidal heating 
$dE/dt\sim (e^2/a_p^6)$ (to second order in $e$, e.g.; \citealt{Batygin+09b}), 
was larger than today. On the other hand, inclusion of tidal migration $\dot{a}_p$
of the planet would have reduced the amount of stellar irradition received at early 
times~\citep{Miller+09}. A proper treatment of both effects is beyond the scope of 
this paper since the orbital eccentricity is not well-constrained \citep{Charb+09}.
Instead, we make the standard assumption that the formation of GJ$\:$1214b yielded 
an initial heat of accretion that was not released before the planet arrived at its 
current location.

The cooling time of our thermal evolution models is the time between the present 
state and the time when the derivative $dT_{\rm eff}/dt$ approaches $-\infty$, 
which corresponds to a hot start that is insensitive to the initial conditions.
While such a treatment of young planets is common (\citealp[e.g.;][]{Baraffe+03}),
it actually ignores the process of planet formation which has been shown to alter
the luminosity during the first tens of millions of years~\citep{Fortney+05}.
This is important for mass determinations from cooling tracks, whereas in our
case it just might induce an error of $\sim 10$ Myrs to the calculated cooling time.

\subparagraph{Love number $k_2$.}
We follow the call by \citet{RagWol09} to tabulate the tidal Love Number 
$k_2$ values of representative models. This quantity is a planetary property 
which solely depends on the internal density distribution.  If known, $k_2$ imposes
an additional constraint on interior structure models. Physically, $k_2$ quantifies 
the quadrupolic gravity field deformation at the surface in response to an external 
perturbing body of mass $M$, which can be the parent star, another planet, or a 
satellite. $M$ causes a tide-raising potential 
$W(r)=\sum_{n=2}W_n=(GM/a)\sum_{n=2}(r/a)^nP_n(\cos\psi)$, where $a$ is the distance 
between the centers of mass (planet and body), $r$ is the radial coordinate of the 
point under consideration inside the planet, $\psi$ is the angle between the planetary 
mass element at $r$ and the center of mass of $M$ at $a$, and $P_n$ are 
Legendre-Polynomials. Each external potential's pole moment $W_n(r)$ induces a change 
$V_n(r)$ in the corresponding degree of the planet's potential, where 
$V_n(R_p)=k_n\,W_n(R_p)$ defines the Love numbers.
For the calculation of $k_2$ we follow the approach described in \citet{ZT78}.

To first order in the expansion of the planet's potential, $k_2$ is proportional to the 
gravitational moment $J_2$ \citep[see e.g.][\sect 4]{Hubb84}. That is why measuring 
$k_2$ will provide a constraint for extrasolar planets that is equivalent to $J_2$ for 
the solar system planets. In particular, $k_2$ is known to be a measure of the 
central condensation of an object, which can be parameterized by the core mass 
within a two-layer model approach. However, as for $J_2$, the inverse problem---the 
deduction of the internal density distribution of the planet from $k_2$, is non-unique.

We stress that the Love number $k_2$ is a \emph{potentially observable parameter} if the 
planet's eccentricity is non-zero. It can be obtained with 
the help of transit light curves that contain information about the tidally induced apsidal 
precession of close-in planets \citep{RagWol09}, or for specific two-planet systems can also be 
derived from measuring the orbital parameters \citep{Batygin+09b}, given that the planets 
are in apsidal alignment and on co-planar orbits \citep{Mardling+10}.

\section{Results}\label{sec:res}

The framework of the three structure classes I-III is used to investigate the possible 
set of models with respect to core mass, internal pressures and temperatures, water to 
rock ratio, and the H/He mass fraction.

\subsection{GJ 1214b with a H/He atmosphere and a rock core}\label{ssec:res_class1}

\begin{figure*}
\epsscale{0.40}
\plotone{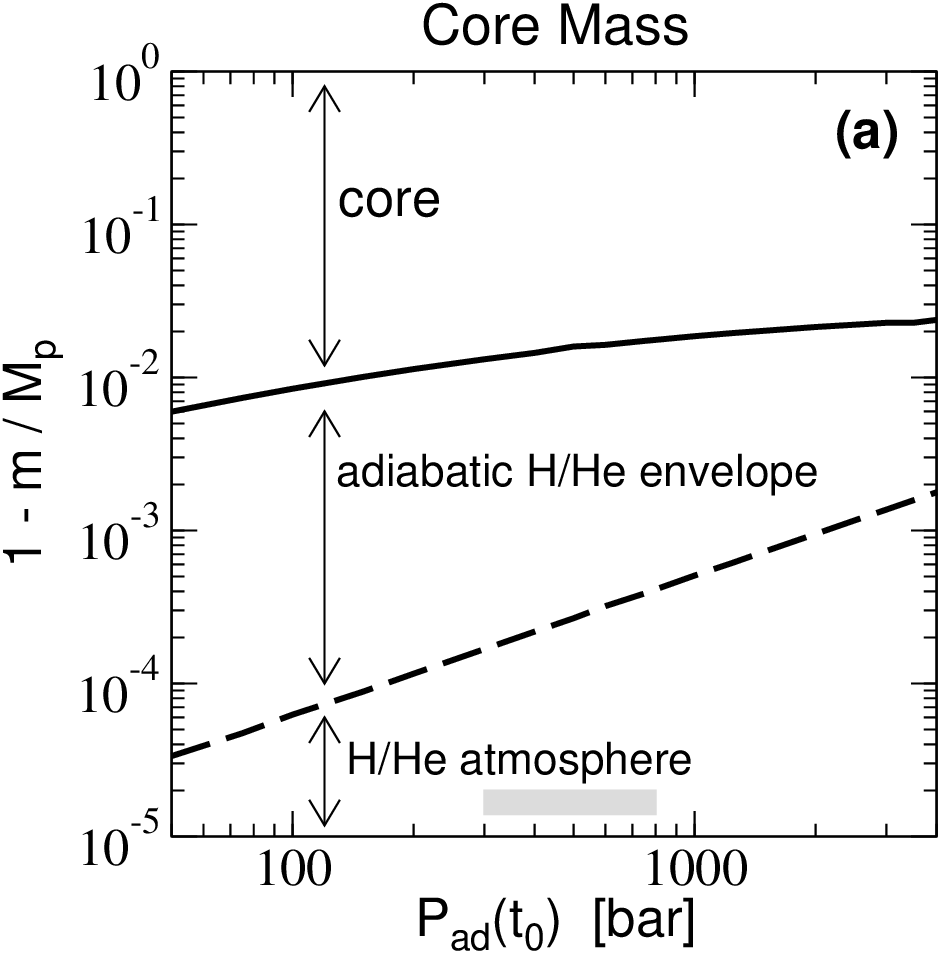}
\plotone{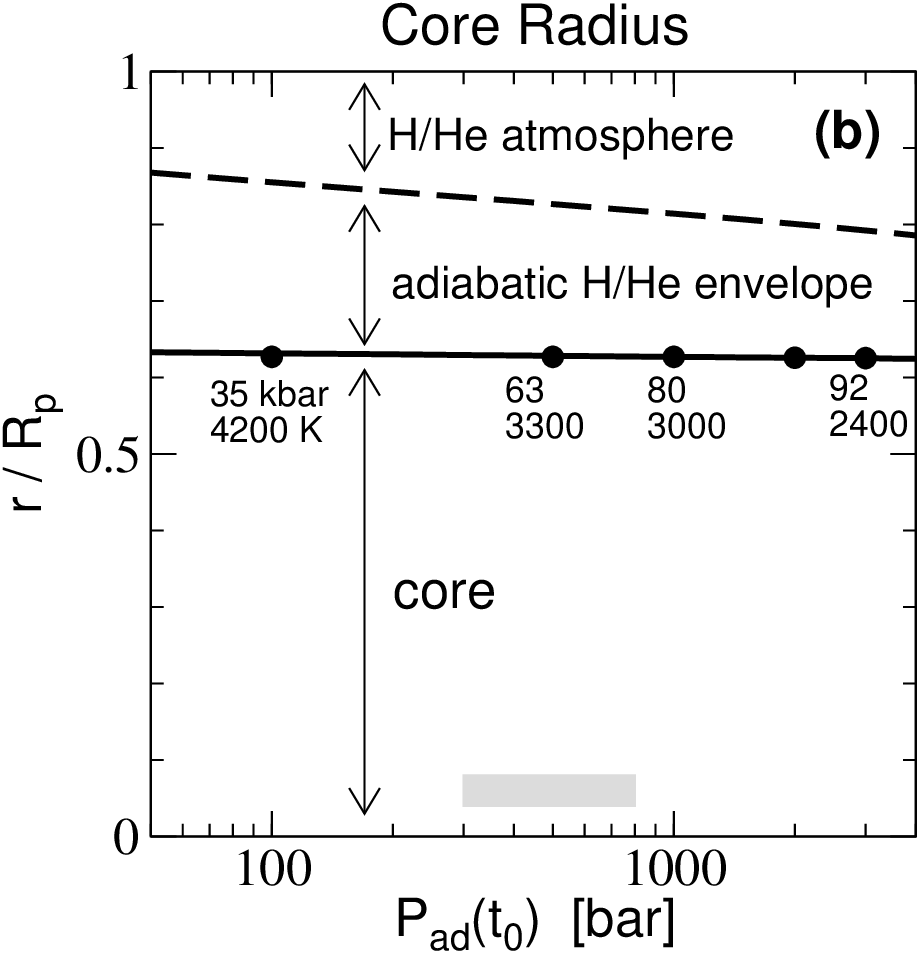}
\epsscale{0.40}
\vspace*{2mm}
\plotone{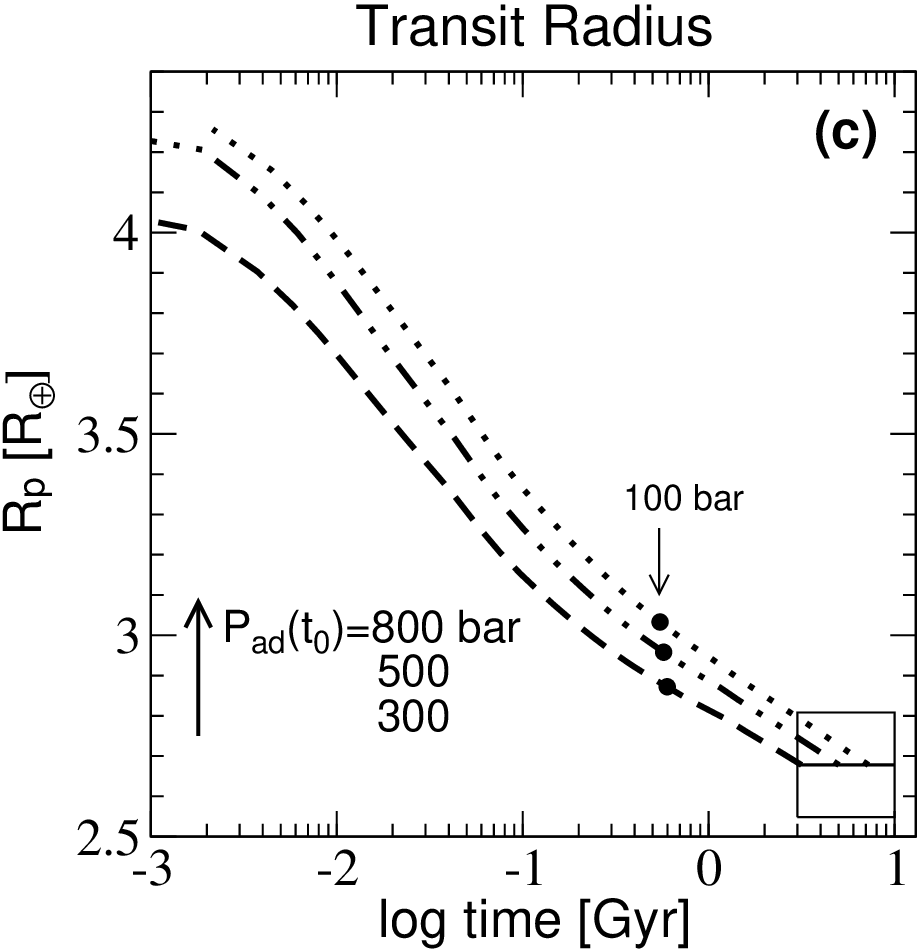}
\plotone{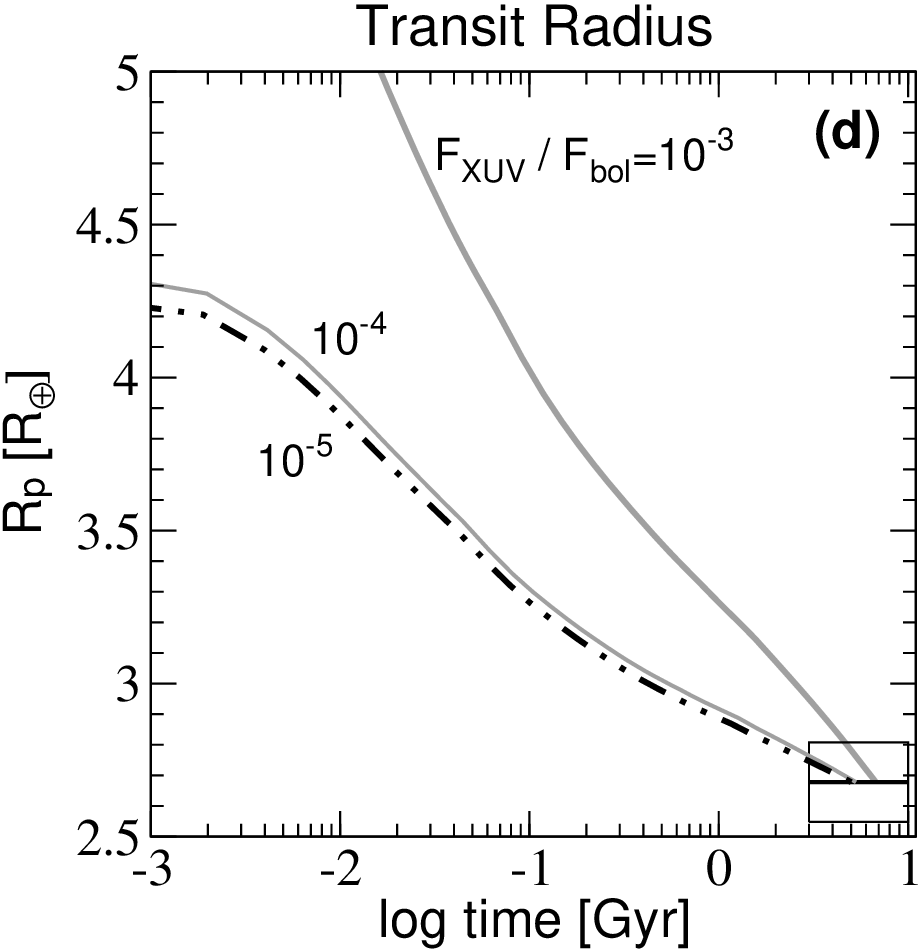}
\caption{\label{fig:HHeR}
Cooling curves and internal structure of models with H/He envelope+rock core. 
All models are adjusted to have $R_{p}(t_0)=2.678\RE$ by the choice of the core mass, 
and $M_{p}(t_0)=6.55\ME$ at present time $t_0$.
\emph{Upper panels:} 
Mass (panel a) and radius (panel b) of the core (\emph{solid line}) and where the 
atmosphere transitions into the adiabatic interior (\emph{dashed}) of present time models 
with $50\leq P_{ad}(t_0)\leq 4000$~bar. Numbers are pressure in kbar and temperature 
in K at the envelope-core boundary for selected models. Single interior models run vertically.
The \emph{gray shaded areas} are a guide to the eye for the allowed $P_{ad}(t_0)$ range that 
is consistent with the thermal evolution.
\emph{Lower panels:} 
Evolution of radius with time. \emph{Panel c:} its dependence on $P_{ad}(t_0$), \emph{panel d:} 
on the insolation as parameterized by $F_{\rm XUV}/F_{\rm bol}$;  \emph{dashed}: model Ia, 
\emph{dotted}: model Ib (see \tab\ref{tab:SM}), \emph{dot-dot-dashed}: an intermediate one 
with $P_{ad}(t_0)=500$~bar. 
These cooling curves are for a specific heat of the core material $c_v=1.0$~J/gK and an 
insolation $F_{\rm XUV}/F_{\rm bol}=10^{-5}$ (\emph{black}), $10^{-4}$ (\emph{thin solid gray}), 
and $10^{-3}$ (\emph{thick solid gray}). \emph{Circles} in panel c indicate the profiles 
during evolution when $P_{ad}(t)=100$~bar. The \emph{boxes} indicate the observational 
error bars of radius (vertical extension) and stellar age (horizontal extension).}
\end{figure*}

\placefigure{fig:HHeR}

For our H/He envelope+rock core models we find a narrow core mass fraction 
range of 0.975$-$0.995 for a wide pressure range $50\leq P_{ad}(t_0)\leq 4000$~bar 
(\fig\ref{fig:HHeR}a). Despite its low mass fraction, the H/He envelope extends over 
$0.37\:R_p$ (\fig\ref{fig:HHeR}b) independent of the envelope mass. This is because 
of its high temperature, which increases from 1030~K at $P_{ad}(t_0)$ to 2400$-$4200~K 
at $\sim 35-90$~kbar at the envelope-core boundary. At these conditions, hydrogen is 
molecular throughout the envelope according to the SCvH-i EOS.  

Not all of the models shown in \fig\ref{fig:HHeR}a,b are consistent with an age
$\tau_\star=3$ to 10 Gyr of the star GJ$\:$1214. If $P_{ad}(t_0)$ decreases (increases), 
internal entropy rises (falls) and the planet will need less (more) time 
to cool down to this state. 
This behavior is illustrated by the cooling curves in Fig.~\ref{fig:HHeR}c, according to 
which an age of 3~Gyr requires $P_{ad}(t_0)\geq 300$~bar; and a much longer cooling time of 
7.2~Gyr is obtained for $P_{ad}(t_0)=800$~bar, whereas $P_{ad}(t_0)=100$~bar would give a 
cooling time below 1~Gyr. With a mass loss rate according to $F_{\rm XUV}/F_{\rm bol}=10^{-5}$
as assumed for the cooling curves in \fig\ref{fig:HHeR}c, a cooling time of 10~Gyr can
not be obtained through a further increase of $P_{ad}$ beyond 800~bar if $L_{radio}(0)$ 
is not to drop below $0.9\times L_{cosmo}$. The colder the interior, the lower $\Tint$ 
as predicted by our model atmsphere grid, and hence the intrinsic energy then can be 
transported through the radiative atmosphere. 
Enhancing $P_{ad}(t_0)$ from 300 to 800~bar lowers $\Tint$ from 42.7 to 31.8~K 
(\tab\ref{tab:SM}). For even colder interiors, the atmosphere is no longer capable of 
radiating away the heat generated by radioactive decay, which would contradict the 
assumption of such a cold interior.

The cooling time increases with the mass loss rate (Fig.\ref{fig:HHeR}d).
For stellar XUV radiation $F_{\rm XUV}/F_{\rm bol}=10^{-5}-10^{-4}$ as typical for quiet 
M-stars~\citep{Scalo+07}, this enhancement is small (see \fig\ref{fig:HHeR}d) and we obtain 
essentially the same range of $P_{ad}(t_0)$ and hence structure models. 
Of  GJ$\:$1214b's initial total mass (initial H/He envelope mass), only 0.0054$-$0.013\% 
(0.41$-$0.74\%) is lost if $F_{\rm XUV}/F_{\rm bol}=10^{-5}$, and about 
0.1\% (6\%) if $F_{\rm XUV}/F_{\rm bol}=10^{-4}$. On the other hand, for a permanent, strong 
irradiation $F_{\rm XUV}/F_{\rm bol}=10^{-3}$ as observed for young, active M-Stars listed 
in the ROSAT catalogue, GJ$\:$ 1214b would have lost 1.8\% (53\%). These numbers are
in agreement with the rough estimates in \sect\ref{ssec:massloss}.
In the last case, existence of a thin H/He atmosphere to-date becomes less likely since 
it begins to require fine-tuning of the initially accreted H/He envelope mass.
We have found cooling tracks with an age of 10~Gyrs or more only if $P\geq 800$~bar and  
the mass loss rate is high, or $L_{radio}(t)\ll L_{cosmo}$, which we do not favor.

Models with $300\leq P_{ad}\leq 800$~bar as constrained by our evolution calculations 
have $M_{core}=6.434-6.464\ME$ and $k_2=0.0170-0.0183$ implying a high degree of central 
condensation. The presence of an iron core could even enhance the central condensation. 
Therefore, although the H/He layer is low in mass, it significantly strengthens the 
property of central condensation compared to a closer to zero-mass atmosphere
planetary object such as the Earth, the theoretical $k_2$ value of which is $\sim 0.3$ 
\citep{Zhang91}.

Class I models are closest to giant planets that formed within the snowline of the 
disk and did not have enough time and/or material in their surrounding to accrete a 
massive H/He envelope. Class I structure models can best be compared to 'case I' 
models by \citet{RogSea10}, where the difference in composition assumptions can be 
reduced to the core (undifferentiated rock core versus differentiated 
iron-silicate-water-ice core in their models), and a slightly different envelope He 
abundance of respectively 0.27 and 0.28.  

Our obtained H/He mass fraction range, 1.3$-$1.8\%, is due to the uncertainty in 
$\Tint$. This range is much smaller than theirs ($9\times 10^{-3}-6.8\%$), which includes 
uncertainties from the 1$\sigma$ errors of $M_p$ and $R_p$ contributing an uncertainty 
of 0.4$-$1.6\%, from their $\Tint$ values used contributing up to 1.5\%, and from
the uncertainty in core composition (pure iron, iron-silicates-water, or pure water).
Since iron is included in our rock EOS, we consider (1.8\%+1.6\%=)~3.4\% a reliable upper 
limit of the H/He mass fraction if observations reveal a low-mean molecular weight 
atmosphere. Recent observations of GJ$\:$1214b's atmosphere in the 0.78 to 1 $\mu$m 
wavelengths range with the VLT facility's UT1 telescope suggest a mean molecular weight 
of 5 g/mol or more and thus disfavor a H/He-dominated atmosphere \citep{Bean+10}. 
On the other hand,  the presence of clouds or hazes in such an atmosphere could mimic
a short scaleheight, hence high mean moelcular weight, and cannot be excluded by current
observations and model atmospheres.

\subsection{GJ 1214b as a water planet with a rock core}\label{ssec:res_class2}

\placefigure{fig:H2OeR}
\begin{figure*}
\epsscale{0.37}
\plotone{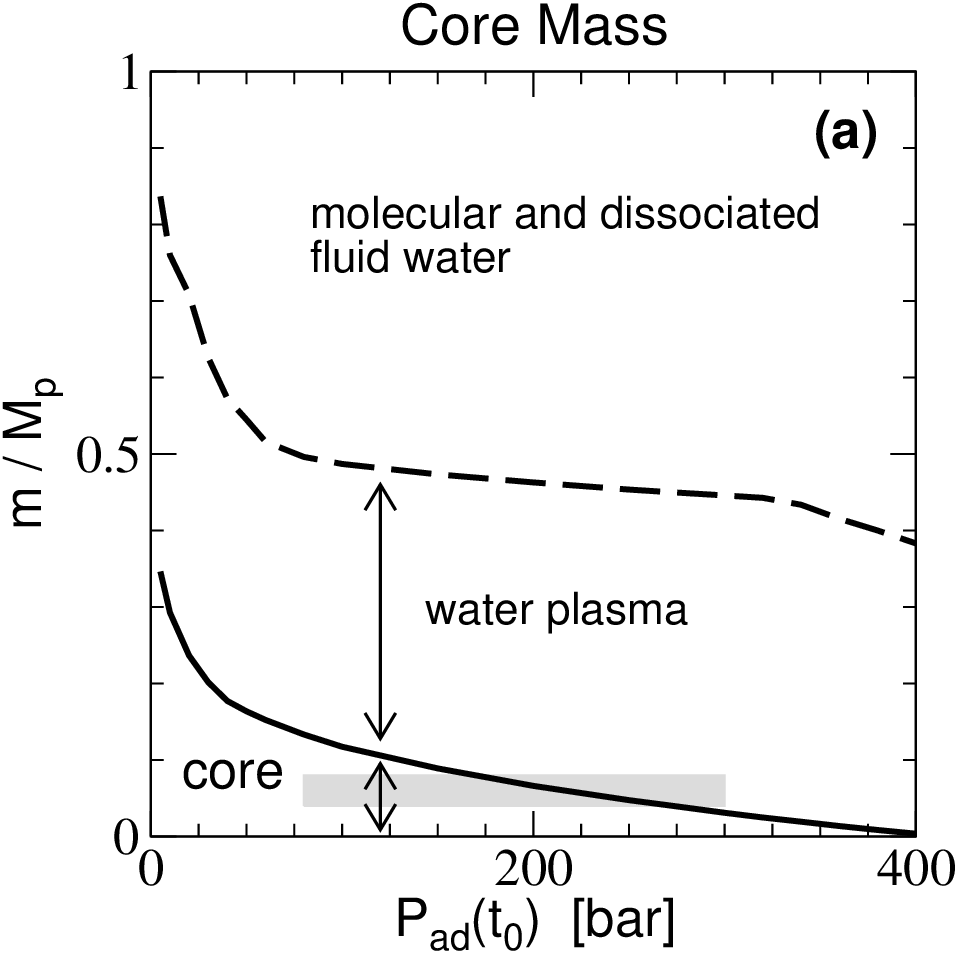}
\plotone{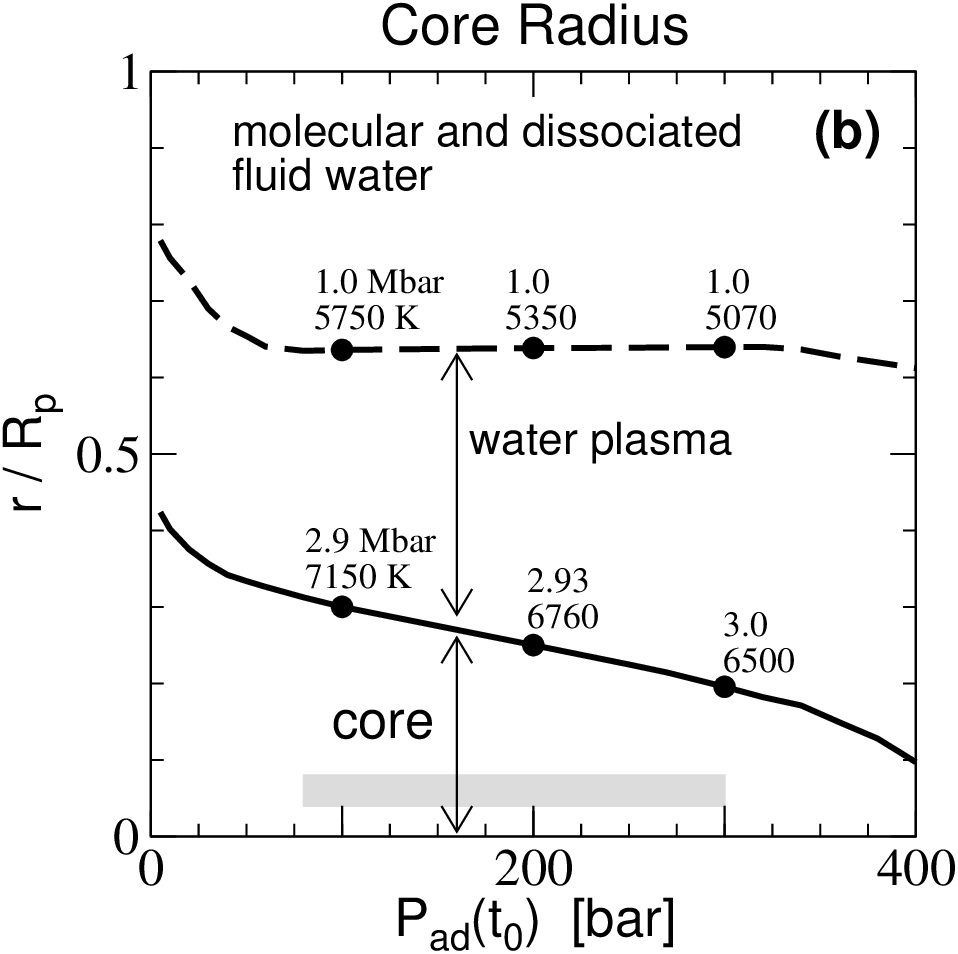}
\epsscale{0.36}
\plotone{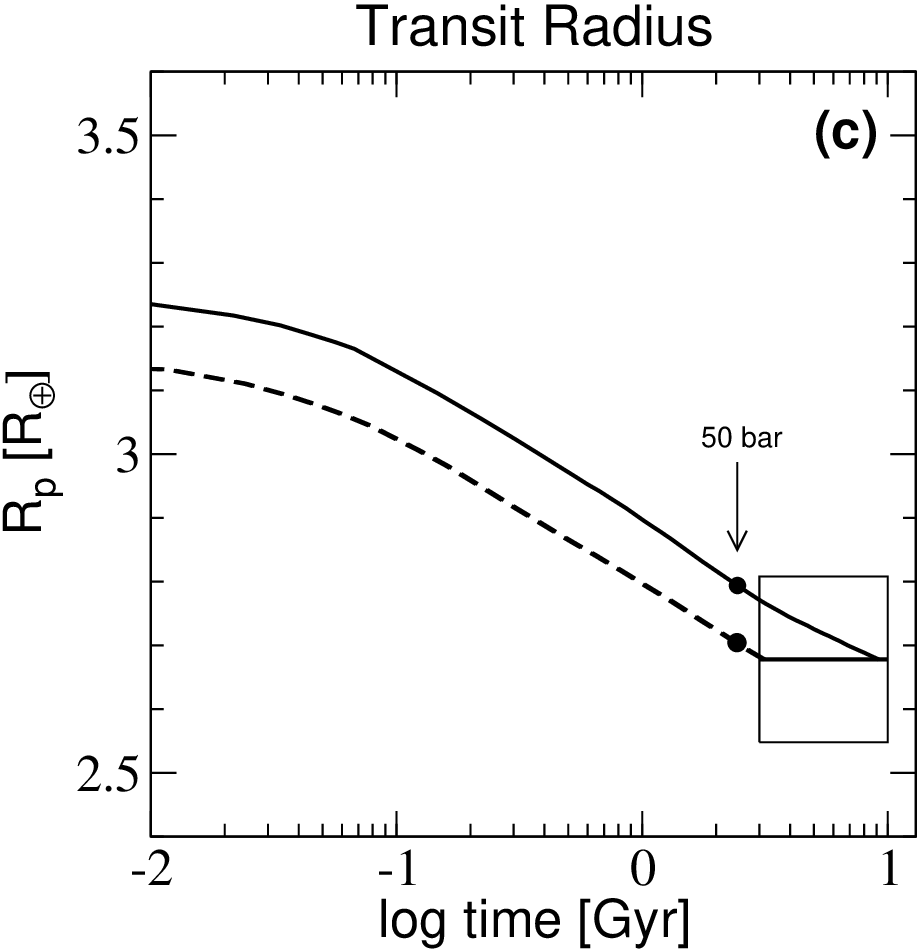}
\caption{\label{fig:H2OR}
Cooling curves and internal structure of models with water envelope+rock core.
All models are adjusted to have $R_{p}(t_0)=2.678\RE$ by the choice of the core mass.
\emph{Left (panel a):} Core mass (\emph{solid line}) and mass shell where water in the 
envelope enters the plasma phase (\emph{dashed line}) of present time structure models for 
transition pressures $5\leq P_{ad}(t_0)\leq 400$~bar. 
Single interior models run vertically. The \emph{gray shaded areas} are a guide to the eye 
for the allowed $P_{ad}(t_0)$ range that is consistent with the thermal evolution (in panel c).
\emph{Middle (panel b):} Same as (a) but radius coordinate. Numbers at models highlighted 
by \emph{filled circles} are pressure in Mbar and temperature in K.
\emph{Right (panel c):} Evolution of radius of structure models IIa (\emph{dashed}) 
and IIb (\emph{solid}), see \tab\ref{tab:SM}. \emph{Circles} indicate those 
profiles during evolution when $P_{ad}(t_0)=50$~bar.
}
\end{figure*}

The derived core mass of water envelope+rock core models responds much more strongly to 
a change of the onset of the adiabatic part of the envelope than models with H/He envelope.
At $P_{ad}(t_0)\approx 400\:$bar, the core mass reaches zero: a pure water planet 
(\fig\ref{fig:H2OR}a). Deeper isothermal regions would only be possible if some amount 
of the pure water planet would be replaced by lighter elements such as methane or ammonia.
Warming up the deep envelope by an outward shifting of the onset of the adiabatic
region below 50~bar is accompanied by a strong rise in core mass. A solar water:rock ratio 
of 2.5 occurs for $P_{ad}=5\:$bar---when the deep interior is extremely hot. 
Smaller ratios would be possible only if the isothermal region is allowed to disappear. 
We next consider whether such models are consistent with the cooling time.   

Because of the relatively small core mass fraction of class II models, varying the specific
heat of the core has a negligible effect on the cooling time. For $80\leq P_{ad}(t_0)
\leq 300\:$bar the resulting cooling times are consistent with $\tau_\star$, see
\tab\ref{tab:SM} and \fig\ref{fig:H2OR}c. 
Those models have $M_{core}=0.20-0.87\:\ME$, and a water:rock ratio of 6.5$-$31. This 
is much higher than the solar ice to rock ratio (I:R)$_{\odot}\sim 2.5$, which would 
give a cooling time shorter than 3~Gyr.

In contrast to models Ia,b, models IIa,b are weakly centrally condensed as 
parametrized by their high $k_2$ values of respectively 0.57 and 0.74.

\begin{figure}
\epsscale{1.0}
\plotone{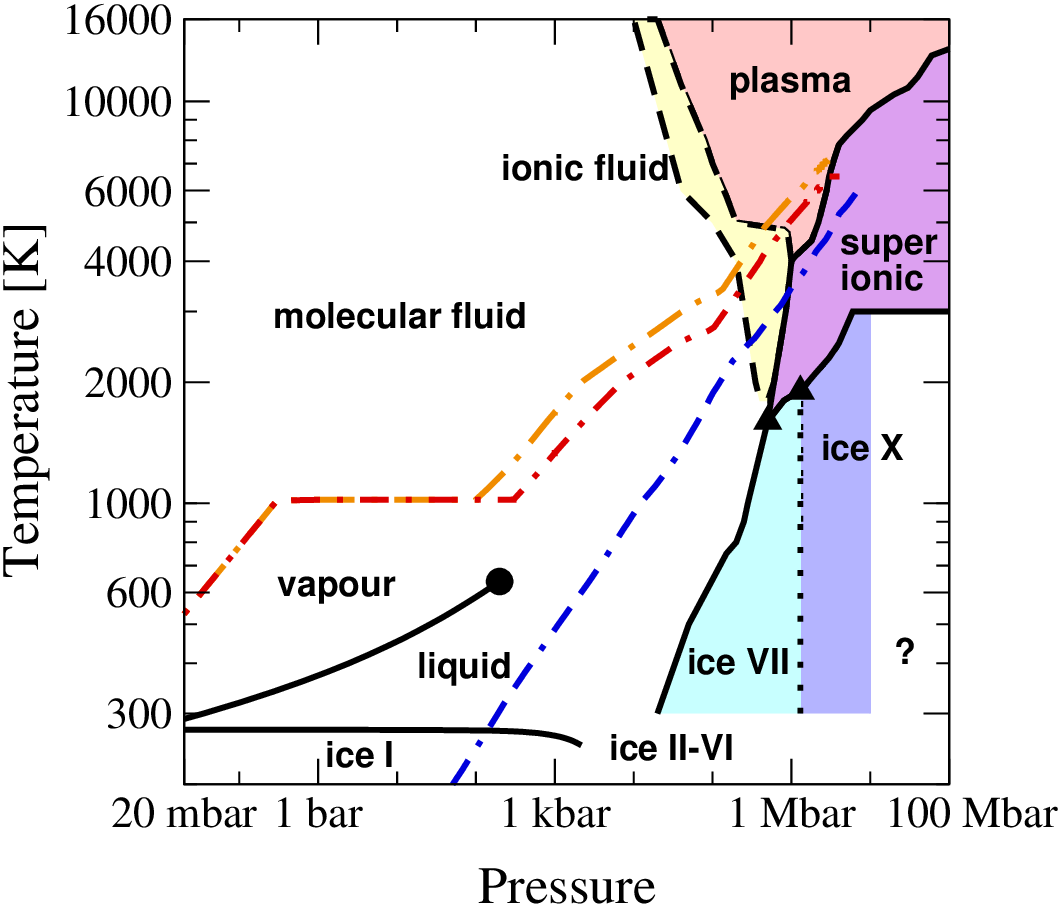}
\caption{\label{fig:wphases}
Water phase diagram for 20~mbar$<P<100$~Mbar and GJ$\:$1214b water envelope+rock core 
models with $P_{ad}(t_0)=100$~bar (\emph{orange dashed-dot}) and 300~bar 
(\emph{red dashed-dot}). The Neptune profile (\emph{blue dashed-dot}) is adopted 
from \citet{Redmer+10}.}  
\end{figure}

Class II models are different from those by \citet{Fu+10} who consider cold water
planets with ice or liquid ocean layers above silicate/iron cores.  
GJ$\:$1214b is not that cold. Our models also differ from the water steam atmosphere models 
of CoRoT-7b by \citet{Valencia+10} who find water envelope + silicate/iron core models 
where water in present CoRoT-7b contributes at most 10\% to the total mass and is in the 
vapor phase or supercritical molecular phase.  They also differ from the water steam 
atmosphere models by \citet{RogSea10} who describe the deep interior by a water-ice 
equation of state, whereas according to the phase diagram (\fig\ref{fig:wphases}) 
(updated from \citealp{French+09,Valencia+10}), water would be in the plasma phase 
in GJ$\:$1214b.

\placefigure{fig:wphases}

In our class II models, water transitions from the vapor phase to supercritical molecular 
water still in the atmosphere, becomes dissociated into an ionic fluid at about 4000$\:$K 
and 0.2~Mbar, and finally fully dissociated and ionized for $T\geq 5100\:$K and $P\geq 1\:$Mbar 
(\fig\ref{fig:H2OR}b) forming a plasma with electronic conductivity $\geq 100\:\rm\:\Omega/cm^2$ 
\citep{Redmer+10}.

Under these circumstances, a planetary 
interior can be able to maintain a dynamo generating a dipolar magnetic field. In Uranus 
and Neptune, the magnetic field may be generated in a thin shell \citep{StanBlox06} of 
possibly ionic water \citep{Nellis+88}. This view of the cold ($\Teff\approx 59$ K) outer 
planets Uranus and Neptune is supported by the Neptune adiabat in \fig\ref{fig:wphases}, 
while in GJ$\:$1214b, since it is warmer, the fluid conductive envelope would extend down 
to the small core, somewhat akin to Jupiter, and therefore preferably lead to a dipolar field, 
according to the field geometry considerations by \citet{StanBlox06}.

\subsection{GJ 1214b as a H/He/H$_2$O planet with a rock core}\label{ssec:res_class3}

This structure type resembles the outer envelope of Uranus and Neptune if $Z_1<0.4$, 
or their inner envelope if $Z_1>0.7$ according to three-layer Uranus and Neptune
models of \citet{FN09}.  A strong enrichment in metals of the outer H/He layer in Uranus 
and Neptune is necessary to match the gravity field data, and also some admixture of light
elements in the deep interior.

\begin{figure*}
\epsscale{0.37}
\plotone{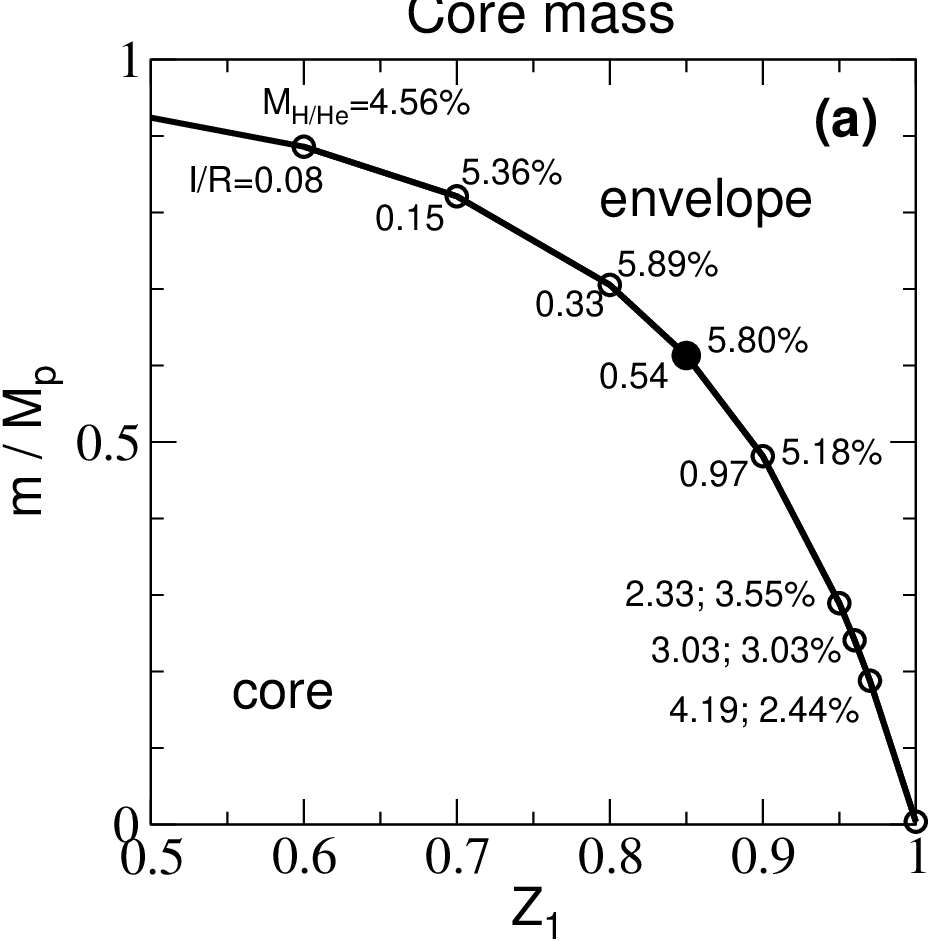}
\plotone{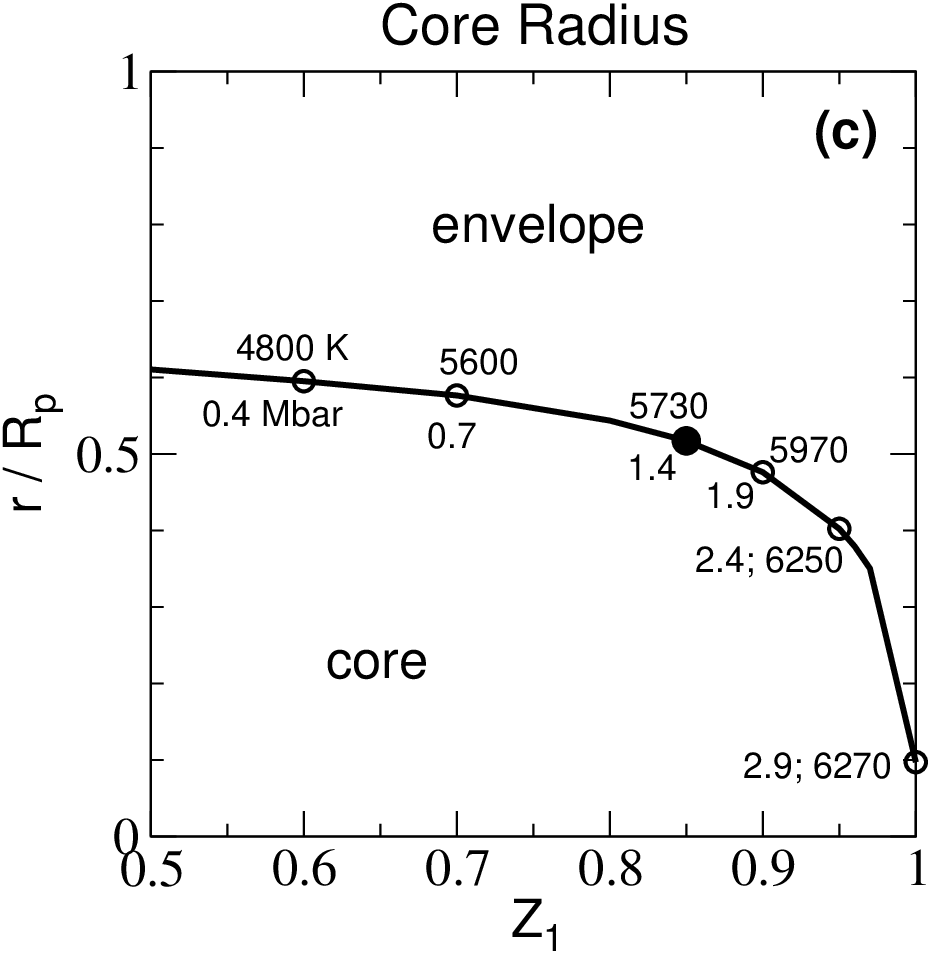}
\epsscale{0.36}
\plotone{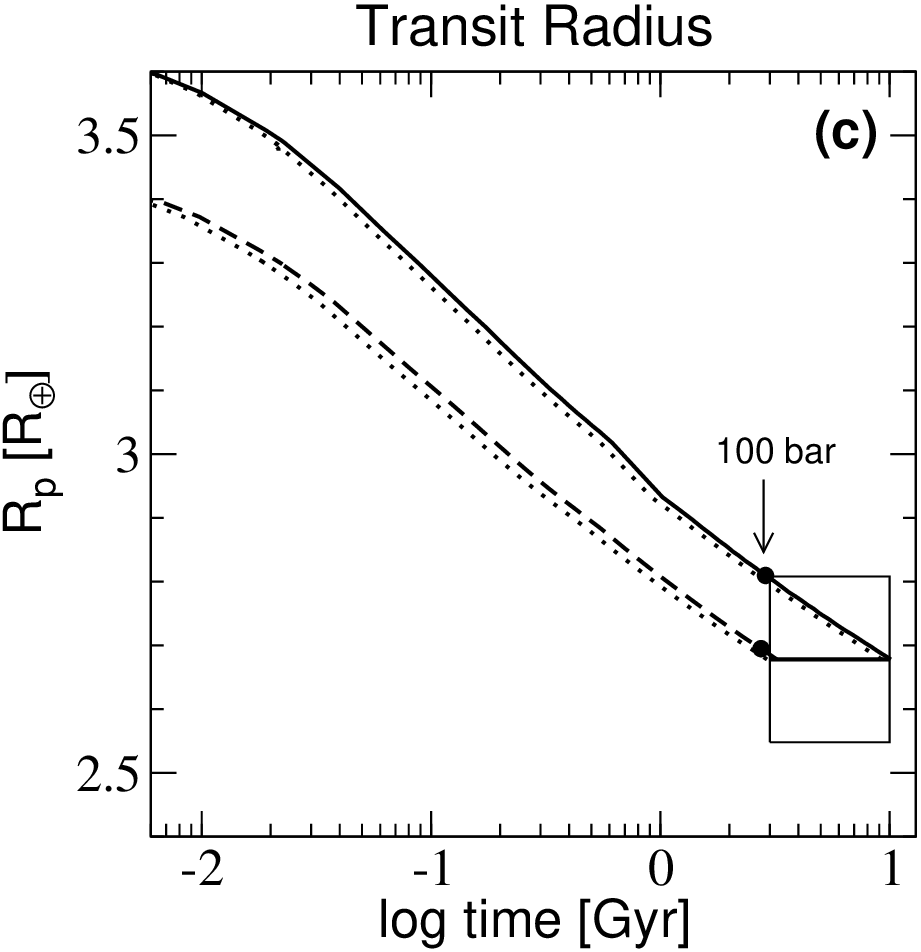}
\caption{\label{fig:HHeH2OR}
Cooling curves and internal structure of H/He/H$_2$O envelope+rock core models 
with various envelope metallicities. 
\emph{Left:} The dependence of the core mass (in $M_p$) on the envelope metallicity $Z_1$ of 
present time structure models with $P_{ad}(t_0)=400$~bar. Numbers at selected models 
are the I:R ratio (\emph{first number}) and the entire planet's H/He mass fraction 
in percent. Model IIIb is highlighted (\emph{filled circle}). 
\emph{Middle:} Core radius in $R_p$. Numbers are pressure in Mbar and temperature 
in K at the envelope-core boundary for the same models as in the left panel.
\emph{Right:} Evolution of radius with time for models IIIa (\emph{dashed}) and IIIb 
(\emph{solid}), see \tab\ref{tab:SM}, using specific heat of the core material 
$c_v=1.0$, or 0.5~J/gK (\emph{dotted}). \emph{Circles} in panel c indicate those profiles 
during evolution when $P_{ad}(t)=100$~bar.
}
\end{figure*}

\placefigure{fig:HHeH2OR}

In \S$\:$\ref{ssec:res_class2} we have seen that water envelope+rock core models of 
GJ$\:$1214b yield supersolar I:R ratios of more than $2.6\times$solar. Lowering 
this ratio can be achieved by replacing water with hydrogen and helium. The two 
limiting cases of this implementation are a structure where a H/He layer 
is on top of a water layer, or a homogeneous mixture of H/He and H$_2$O.
We find that in the first case, such differentiated three-layer models (H/He, water, rock) 
can not have $1\times$(I:R)$_{\odot}$ \emph{and} be in agreement with $\tau_\star$, 
the reason of which is the following.
Class I and class II models require a radiative atmosphere down to 80 bar or more at present 
in order to meet $\tau_\star$. If composed solely of H/He, the atmosphere extends over 
about $0.4\:\RE$. In order to match a remaining radius $r(M_p)\approx 2.3\:\RE$, the core 
mass fraction of the remaining water+core body is of the order of 20-50\% 
\citep[see][\fig 8]{Valencia+10}. We find I:R$<0.9$ and $M_{core}=3.5-4.2\:\ME$ for 
this case of differentiated models. Increasing $P_{ad}$ increases the depth of the thin 
H/He atmosphere, thereby lowering the I:R ratio even more. Increasing the planet's mean 
density within the $1\sigma$ error bars of $M_p$ and $R_p$ allows for I:R up to at most 
$0.56\times$(I:R)$_{\odot}$.

Consequently, the only way to obtain a solar I:R ratio is to limit the radius of 
the H/He atmosphere by enhancing its mean molecular weight \citep{MillerRicci+09} 
through admixture of water. Here we consider the case of equal metallicity in the 
radiative atmosphere and in the adiabatic envelope (our class III models) 
as parameterized by the water mass fraction $Z_1$. Figure \ref{fig:HHeH2OR}a shows  
the change of the I:R ratio of single models. The water to core mass ratio rises 
moderately up to $Z_1=0.9$, passes $1\times$(I:R)$_{\odot}$ at $Z_1=0.95$, 
and then rises rapidly up to the values found for class II models. This behavior 
depends very weakly on the choice of $P_{ad}(t_0)$.

For class III models, the resulting H/He mass fraction of the planet
(see \fig\ref{fig:HHeH2OR}a) is about 2$-$3 times larger than in case of a H/He layer 
on top of a water layer. It reaches the maximum for $Z_1\approx 0.80$ (the core mass must 
not be too large, requiring a high metallicity, and also $Z_1$ not too close to 1).
We find a planetary H/He mass fraction $\leq 5.9$\% if $P_{ad}(t_0)=400\:$ bar as in 
\fig\ref{fig:HHeH2OR}a,b, and slowly rising with $P_{ad}(t_0)$ up to 7\% (colder envelopes 
reduce the core mass). However, a colder present time interior would take longer than 10~Gyrs 
to cool. For the cooling curve calculations we choose a metallicity $Z_1=0.85$, which gives a 
H/He mass fraction close to the maximum value but also a core mass below $2/3\:M_p$, hence a 
real alternative to classes I and II. Figure \ref{fig:HHeH2OR}c shows that the isothermal 
region of present GJ$\:$1214b must end between 120 and 400~bar to give consistency with a 
cooling time of 3 to 10~Gyrs.  

With $Z_1=0.85$, $T_{core}=5730\:$K, and $P_{core}=1.4\:$Mbar (\fig\ref{fig:HHeH2OR}b), 
model IIIb resembles the interior of Uranus and Neptune in composition and 
temperature \citep{FN09}. Lower in total mass, the pressure does not rise up 
to 5$-$7~Mbar as in the outer solar system giant planets, so that water will not adopt 
the superionic phase according to the phase diagram of water, but remain in a fluid 
state in GJ$\:$1214b (\fig\ref{fig:wphases}). This property bolsters our 
assumption of a homogeneous mixture of water with hydrogen and helium.

\section{Discussion}\label{sec:discuss}

\subsection{Structure assumptions}

Our GJ$\:$1214b interior models rely on a separation of the interior into a rock core 
and one homogeneous envelope of the same composition as in the visible atmosphere. 
In contrast, giant and terrestrial planets in the solar system are not
successfully described by such a two-layer structure but require the assumption
of various internal layer boundaries to be consistent with the atmospheric
He abundance and the gravity field data (giant planets: see 
\citealp{GudZha99,SauGui04,N-Jupiter+08,FN09}), long-term spacecraft tracking 
data (Mars: see \citealp{Konopliv+06}), and with seismic data (Earth).

\subsubsection{class III: H/He phase separation?}

For the giant planets, a layered structure is suggested in part because of a measured 
atmospheric depletion in He indicating H/He phase separation, and in part because of
relatively low measured $J_4/J_2$ ratios indicating an enhancement with 
metals in the deep interior. According to experimental and theoretical 
data on H/He demixing (see e.g.~\citealp{Lorenzen+09,Morales+09}), immiscibility 
of He in H might also occur in our class III models close to the core-mantle 
boundary. Given the low interior temperatures (relative to Saturn) this could very strongly 
deplete most of the envelope in helium \citep{FH04}.  While accurate measurements of the 
atmospheric He abundances are extraordinary difficult to perform without the use of entry 
probes even in case of solar giant planets, this effect can affect the current depth 
of the isothermal region (120$-$400~bar) that we have derived from the cooling time calculations.  
This would be a completely different pathway towards a helium-depleted hydrogen atmosphere 
than the outgassing scenario discussed in \citet{RogSea10}. On the other hand if $Z_1\ll 0.85$,
then the envelope does not become massive and dense enough for He sedimentation to occur.

\subsubsection{class II: Incomplete Differentiation?}

Our class II models require a process that causes a downward sedimentation 
of rocks in order to separate out a water layer, as in Ganymede \citep{KirkStev87}. 
Up to now, there is no experimental evidence that water and silicates or iron become 
immiscible under high pressure, and the timescale for gravitational settling 
before the onset of convection, terminating gravitational settling, is essentially 
unknown for Ganymede \citep{Kimura+09}. 
Since the larger primordial heat deposited in the $\sim 260$ times more massive 
GJ$\:$1214b might have caused a rapid onset of convection, a water+rock interior 
of GJ$\:$1214b is possibly not fully differentiated. Internal layer boundaries 
dividing regions with different water to rock ratios can not be excluded, and the
core rather be an ice-rock mixture as suggested for Callisto \citep{Nagel+04} 
than pure rocks. On the other hand, more detailed envelope models of GJ$\:$1214b 
would be underdetermined by current observational parameters.

\subsubsection{class I: Choked off giant planet formation?}

The core mass of our class I models is within the range of rock core masses
currently proposed for Jupiter \citep{FN09}. With an initial H/He atmosphere of only 
1$-$2\% $M_p$, GJ$\:$1214b appears to be a giant planet whose envelope mass accretion was
choked off during formation. It is of general interest for more super-Earth planets to be
detected to see whether this is a common outcome of planet formation.

\subsection{Composition}

From interior and atmosphere models that are consistent with the observationally
derived parameters $\Teq$, $M_p$, and $R_p$ after 3$-$10 Gyrs of cooling, 
the most certain conclusion we can draw about the composition of GJ$\:$1214b is
a metallicity of 94$-$100\%. The lower limit can further shrink somewhat if the planet 
in reality is dry, and the mass fraction of water in our class III models then 
resembles a mixture of H/He and rocks. In contrast, the mass fractions 
of water, used as a proxy for the ices H$_2$O, CH$_4$, NH$_3$, and H$_2$S, is
essentially unconstrained (0$-$97\%) as is the mass fraction of rocks (3$-$99\%).

Two further observables we can hope to attain in the near future are the Love number
$k_2$ from transit timing variations or from the shape of the transit light curves 
\citep{RagWol09}, and second the mean molecular weight in the atmosphere from 
transmission spectroscopy, in particular from the wavelength dependence of star light 
absorption in the planetary atmosphere during transit \citep{MillerRicciFor10}. 

\subsubsection{Love number $k_2$}

\begin{figure}
\epsscale{0.95}
\plotone{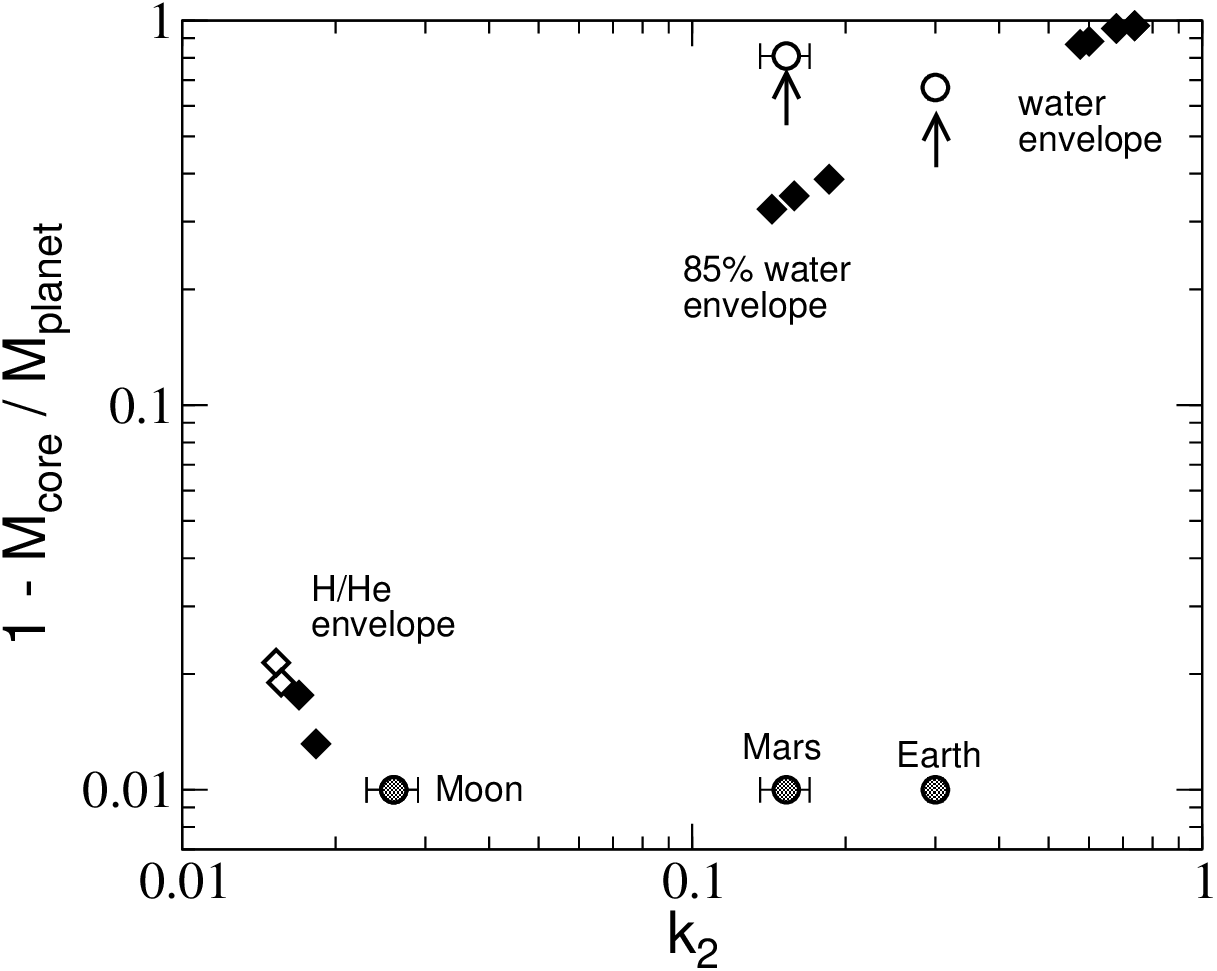}
\caption{\label{fig:Mck2}
Collection of $\{M_{core},k_2\}$ pairs obtained for class I-III models of GJ$\:$1214b
(\emph{diamonds}), that meet the age constraint (\emph{filled}) or not (\emph{open});
and measured $k_2$ values with error bars of Moon, Mars, and Earth (M~M~E) (\emph{circles}).
According to the definition of $M_{core}$ used for GJ$\:$1214b, M~M~E would have 
$M_{core}>0.999999M_p$, which we placed at $0.99M_p$. Considering alternatively $M_{core}$ 
as the iron core of these objects shifts the \emph{filled circles} to the 
\emph{open circles}. This figure demonstrates degeneracy of $k_2$ in two-layer 
models with thin atmospheres.
}
\end{figure}

The $k_2$ values of models from our classes I-III differ
greatly from each other and thus we consider $k_2$ a useful quantity
to discriminate between atmospheres of different mean molecular weight.
The trend of decreasing $k_2$ value with increasing core mass that is
known for two-layer models of Jupiter-mass giant planets with a core and one 
envelope \citep{Batygin+09b}, or for n=1 polytropic planets in general
\citep{Kramm+10}, is confirmed by our two-layer models of GJ$\:$1214b. An 
illustration of the $k_2(M_{core})$ behavior is shown in \fig\ref{fig:Mck2}, 
which contains the same data as \tab\ref{tab:SM} and also some intermediate points as
well as two solutions of the H/He envelope models with respectively $P_{ad}(t_0)=1$ 
and 2~kbar, which are too cold as explained in \sect\ref{ssec:res_class1}. 

\placefigure{fig:Mck2}

However, when the H/He atmosphere becomes thin and its mass low enough
($<2\%$), this planet begins to more closely resemble a relatively homogeneous
rock body than a core+envelope planet, and $k_2$ rises again.
This degeneracy of $k_2$ that is otherwise well-known for multi-layer models is 
--at first glance-- a surprising finding in \emph{two}-layer super-Earth models. 
As \fig\ref{fig:Mck2} suggests, pure H atmosphere models of GJ$\:$1214b 
\citep{RogSea10} would approach the $k_2$ values of solar system bodies 
such as the Moon, i.e.~\emph{higher} than that of H/He atmospheres. 

Moon, Mars, and Earth can be considered as remnant protoplanetary cores ($M_{core}=M_p$)
that did not accumulate sufficient mass to accrete a significant envelope.
On the other hand, because measured $k_2$ values of Earth \citep[see][]{Ray+01}, Mars 
\citep{Yoder+03,Konopliv+06}, and Moon \citep{Zhang92} cannot be explained by a 
homogeneous rock interior but require the assumption of a dense, iron-rich core 
\citep[e.g.][]{ZGM09}, they can also be considered as objects with small 
(iron) core and large (silicate) mantle ($M_{core}\ll M_p$). This places Mars and Earth 
close to our GJ$\:$1214b models with H/He/Z envelope in the $k_2-M_{core}$ diagram, 
implying that $k_2$ is a degenerate quantity with respect to composition, too.
Separation into layers of different composition owing to phase differentiation 
and phase transitions in rocks \citep{Valencia+07a} has likely also occured in 
GJ$\:$1214b. Such an advanced treatment of a core would enhance the level of 
central condensation, and result in even lower minimum $k_2$ values below 0.017.
The same trend is expected for inclusion of solid-body effects, which will be important
for colder super-Earth planets and those with less extended, less massive atmosphere.
Theoretical $k_2$ values for the terrestrial objects are in good agreement with 
the observations when making the simplifying assumption 
of an elastic interior \citep{Zhang92,Yoder+03} (in which case the shear modulus 
becomes frequency-independent, i.e.~$k_2$ a static Love number), while their observed 
moments of inertia are close to 0.4 indicating a nearly homogeneous interior,
the theoretical Love number $k_2$ of which would approach 1.5 if the body were
fluid and compressible.

\subsubsection{Mean molecular weight}

\begin{figure}
\epsscale{0.95}
\plotone{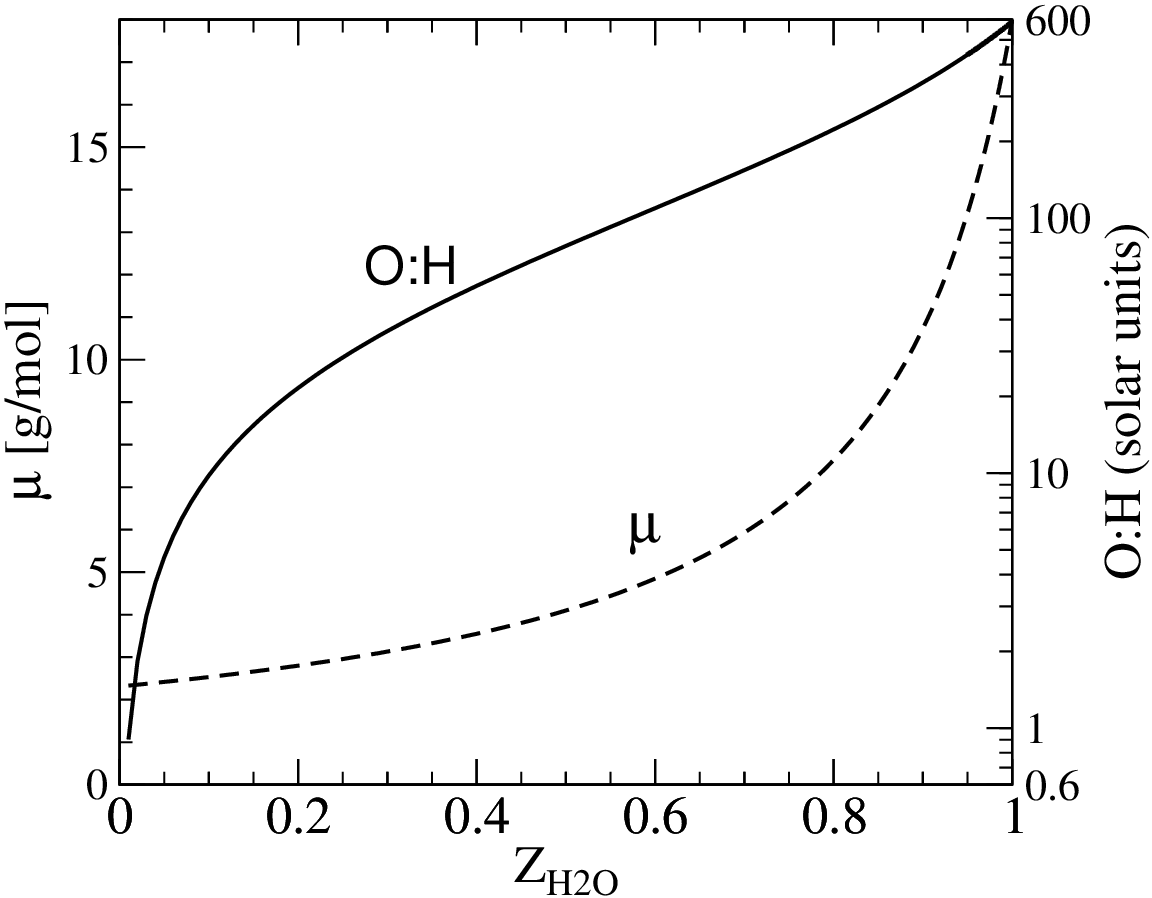}
\caption{\label{fig:mueOH}
Mean molecular weight (\emph{dashed}) and O:H ratio (\emph{solid}) of H/He/H$_2$O mixtures
with water mass fraction $Z$ and Y=0.27. }
\end{figure}

Class III models with $1\times$solar I:R ratio are not excluded by our thermal evolution
calculations. Those models require $Z_1=0.95$, corresponding to a mean molecular weight  
$\mu=13.4\rm\,g\,mol^{-1}$ and 440$\times$solar O:H particle number ratio
(assuming a solar O:H of $0.85\times10^{-3}$ according to \citealt{AndGrev89}).
However, we do not consider such models a realistic description of GJ$\:$1214b.
For rapid translations between metallicity and atmospheric $\mu$ values from 
transmission spectroscopy observations, we present the simple relations between 
$\mu$, O:H, and $Z_1$ in \fig\ref{fig:mueOH}.

\placefigure{fig:mueOH}

In particular, our class III models with $Z_1=0.85$ have $\mu=8.9\rm\,g\,mol^{-1}$, 
a $270\times$ solar atmospheric O:H ratio, 0.2 times solar bulk I:R ratio, and 
$\approx 5.8\%$ of $M_p$ is H/He. For $Z_1<0.55$, our class III models become severely 
rock-dominated with $M_{core}>0.9M_p$, $\mu<4.4\rm\,g\,mol^{-1}$ and O:H~$<92\times$solar. 
A 50$\times$solar O:H ratio as of our model atmosphere grid applied to the thermal 
evolution calculations (see \sect\ref{ssec:atm}) implies $Z_1=0.38$. 
For $\mu=5\rm\,g\,mol^{-1}$ as suggested from the first transmission spectrum
observations by \citet{Bean+10}, \fig\ref{fig:mueOH} gives $Z_{1}=0.7$ and according
to \fig\ref{fig:HHeH2OR}, GJ$\:$1214b could be mostly rocky with an I:R ratio
of 0.15, where the H/He/H$_2$O atmosphere contributes $\sim 18\%$ to the planet's mass
and 40\% to its radius.

\section{Conclusions}

Our results for the composition of the super-Earth mass planet GJ$\:$1214b confirm
that it has a gaseous atmosphere atop a fluid envelope. We find a minimal
total H/He mass fraction of 1.3\% for pure H/He envelopes, which can rise to 
5$-$6\% if the envelope contains 60$-$90\% H$_2$O in mass, and even further 
if also silicates and iron are mixed into the envelope.

Water in GJ$\:$1214b does not solidify within 10~Gyrs of cooling, and it not a liquid 
as is found on Earth's surface, but becomes a plasma if its abundance is high ($Z_1>0.8$).  
This leads to a large water mass ($>1/3 M_p$), where the deep internal matter is warm 
($T>5500$ K) and dense ($P>1$ Mbar). 

The intrinsic heat loss of GJ$\:$1214b after 3$-$10 Gyr of cooling corresponds to
$\Tint=32-52$~K, where the lower bound depends on assumptions about the heat production
by radioactive elements and slightly on mass loss. By our self-consistent mass-loss 
calculations we conclude that GJ$\:$1214b is a genuine super-Earth that has lost an 
insignificant amount of its initial mass, unlike CoRoT-7b.

Nevertheless, some of our computed models may fall into the realm of the unlikely.  
These include class I models that have lost more than 50\% of their initial H/He envelope 
(i.e.~$2\%M_p$ lost) due to extraordinary XUV irradiation and thus require fine-tuning of 
their initial H/He mass fraction. Moreover, extremely low-mass H/He atmospheres in general 
might be an unlikely outcome of planet formation.
The unlikelihood of forming a massive planet with an extreme ice-to-rock ratio of more 
than 6 casts our class II (``water world'') models in doubt.  
We instead favor what is thought to be the most Uranus- and Neptune-like planet models, 
class III. It includes envelopes with $\sim 85\%$ water by mass mixed into H and He 
atop a rock core with about $0.2\times$ solar bulk ice:rock ratio. 
These models have higher H/He masses (5.8\% of the planet mass) than class I models with 
pure H/He atmospheres, lessening worries about the survival of a H/He envelope in the 
face of mass loss.  A general outcome of these models is $\Tint=42\pm 6$~K, 
while the conditions at the core-mantle boundary are $P\approx 1$~Mbar and $T\approx 5700$~K.
The atmosphere can be strongly depleted in helium due to H/He phase separation deeper 
inside the planet (if $Z_1\sim 0.85$) or not be depleted in He if $Z_1$ is much smaller 
than this.

Calculated Love numbers of water envelope models that are consistent with 
the observables $M_p$, $R_p$, $\Teq$, and the age, are in the range
$k_2=$ 0.58$-$0.74, whereas pure H/He envelope models have $k_2\sim 0.018$,
and our favorite models $k_2\sim 0.15$. 
An observational determination of  $k_2$ and the atmospheric mean molecular weight
is crucial for determining the envelope metallicity and the core mass of this 
and other planets that are located along the $M-R$ relation of water planets.
However, ambiguities in composition are an inherent property of such planets.

\acknowledgments{
We thank E.~Kempton (formerly Miller-Ricci) for discussion and the delivery of 
her solar composition and water model atmospheres, T.~Guillot and D.~Valencia 
for valuable comments on the radioactive luminosity of rock material, and
F.~Sohl for helpful discussions on dynamic Love numbers.
This work was supported by the NASA, under grants NNX08AU31G and NNX09AC22G, 
and DFG under grants RE 882/11 and RE 882/12.}

\bibliographystyle{apj}
\bibliography{ms2-refs}

\end{document}